nucl-ex/9511001 3 Nov 1995

# Bounds on new Majoron models from the Heidelberg–Moscow–Experiment*

J. Bockholt, M. Günther, J. Hellmig, G. Heusser, M. Hirsch, H.V. Klapdor–Kleingrothaus †, B. Maier, H. Päs, F. Petry, Y. Ramachers, H. Strecker, M. Völlinger

*Max–Planck–Institut für Kernphysik,*
*P.O.Box 10 39 80, D–69029 Heidelberg, Germany*

A. Balysh, S.T. Belyaev †, A. Demehin, A. Gurov, I. Kondratenko, D. Kotel'nikov, V.I. Lebedev

*Russian Science Center Kurchatov Institute, 123 182 Moscow, Russia*

A. Müller

*Istituto Nazionale di Fisica Nucleare, I–67010 Assergi, Italy*

† *spokesmen of the collaboration*

ABSTRACT

In recent years several new Majoron models were invented to avoid shortcomings of the classical models while leading to observable decay rates in double beta experiments. We give the first experimental half life bounds on double beta decays with new Majoron emission and derive bounds on the effective neutrino–Majoron couplings from the data of the $^{76}Ge$ HEIDELBERG–MOSCOW experiment. While stringent half life limits for all decay modes and the coupling constants of the classical models were obtained, small matrix elements and phase space integrals[1,2] result in much weaker limits on the effective coupling constants of the new Majoron models.

In many theories of physics beyond the standard model neutrinoless double beta decays can occur with the emission of new bosons, so–called Majorons.[3–6] While neutrinoless double beta decay yields the most stringent limits on Majorana masses of neutrinos,[7] the half life bound for Majoron emitting modes yields limits on the effective Majoron–neutrino coupling:

$$2n \rightarrow 2p + 2e^- + \phi \tag{1}$$

$$2n \rightarrow 2p + 2e^- + 2\phi \tag{2}$$

In the classical Majoron model invented by Gelmini and Roncadelli in '81,[4] the Majoron is the Nambu–Goldstone boson associated with the spontaneous breaking of

*submitted to Phys. Rev. Lett.

the $B-L$–symmetry and so generates Majorana masses of neutrinos. As pointed out by Georgi et al.,[5] a sizeable contribution to double beta decay via eq. (1) is expected for the Gelmini–Roncadelli Majoron. However, in this model the Majoron is an electroweak isospin triplet and therefore should contribute the equivalent of two neutrino species to the width of the $Z^0$, which was ruled out by LEP.[8] Also the doublet Majoron[9] was ruled out by this measurement.

On the other hand, Majoron models in which the Majoron is an electroweak isospin singlet[3,10] are still viable. The drawback of the singlet Majoron model is that in these models the Majoron couples to the neutrino at tree level with a coupling strength of roughly $g \simeq (m_{\nu_L}/v_{BL})$, where $v_{BL}$ is the symmetry breaking scale. In order to preserve existing bounds on neutrino masses and at the same time get an observable rate for Majoron emitting double beta decays the singlet Majoron model requires severe finetuning.

To avoid such an unnatural finetuning in recent years several new Majoron models have been constructed, where the terminus Majoron means in a more common sense light or massless bosons with couplings to neutrinos. Since all these models were invented with the same intention of giving observable contributions to double beta decays, we felt motivated to analyze the experimental data on $^{76}$Ge to determine the experimentally allowed size of the effect.

The main novel features of the "New Majorons" (fig. 1 and fig. 2) are that they are not restricted to Goldstone bosons breaking a global lepton number symmetry. Majorons carrying leptonic charge appear in models, where the Majoron is responsible for breaking down an extended symmetry group to the global lepton number symmetry.[11] In vector Majoron models one assumes this extended group to be gauged and the Majoron becomes the longitudinal component of a massive gauge boson[13] emitted in double beta processes. For simplicity we will call it Majoron, too. Also Majorons which are no Goldstone bosons[11] are possible and decays with the emission of two Majorons can occur in models with Majoron fields carrying one unit of lepton number.[12] The latter is mediated by a sterile neutrino.

In tab. 1 the nine Majoron models we considered are summarized.[12,13] It is divided in the sections I for lepton number breaking and II for lepton number conserving models. The table shows also whether the corresponding double beta decay is accompanied by the emission of one or two Majorons.

The next three entries list the main features of the models: The third column lists whether the Majoron is a Goldstone boson or not (or a gauge boson in case of vector Majorons, denoted model IIF). In column four the leptonic charge L is given. In column five the "spectral index" $n$ of the sum energy of the emitted electrons (fig. 3) is listed. (The spectral index is defined from the phase space of the emitted particles, $G \sim (Q_{\beta\beta}-T)^n$, where $Q_{\beta\beta}$ is the Q–value of the decay and $T$ the sum energy of the two electrons.) From experimentators point of view the nine considered models can be reduced to these three different spectral shapes ($n = 1, 3, 7$) and the two neutrino emitting decay with $n = 5$ (fig. 3).

| case | modus | Goldstone boson | L | n |
|---|---|---|---|---|
| IB | $\beta\beta\phi$ | no | / | 1 |
| IC | $\beta\beta\phi$ | yes | / | 1 |
| ID | $\beta\beta\phi\phi$ | no | / | 3 |
| IE | $\beta\beta\phi\phi$ | yes | / | 3 |
| IIB | $\beta\beta\phi$ | no | -2 | 1 |
| IIC | $\beta\beta\phi$ | yes | -2 | 3 |
| IID | $\beta\beta\phi\phi$ | no | -1 | 3 |
| IIE | $\beta\beta\phi\phi$ | yes | -1 | 7 |
| IIF | $\beta\beta\phi$ | Gauge boson | -2 | 3 |

Table 1. Different Majoron models according to Bamert/Burgess/Mohapatra[12]. The case IIF corresponds to the model of Carone[13].

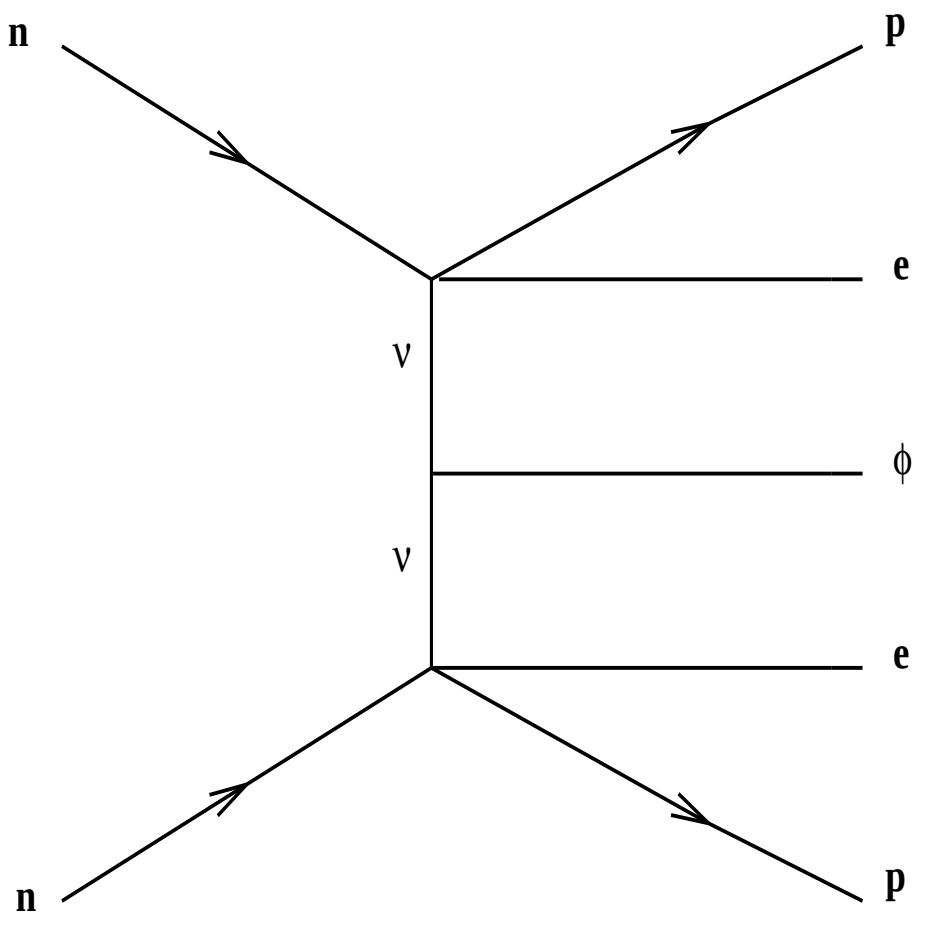

Fig. 1. Feynman graph for $\beta\beta\phi$–decays

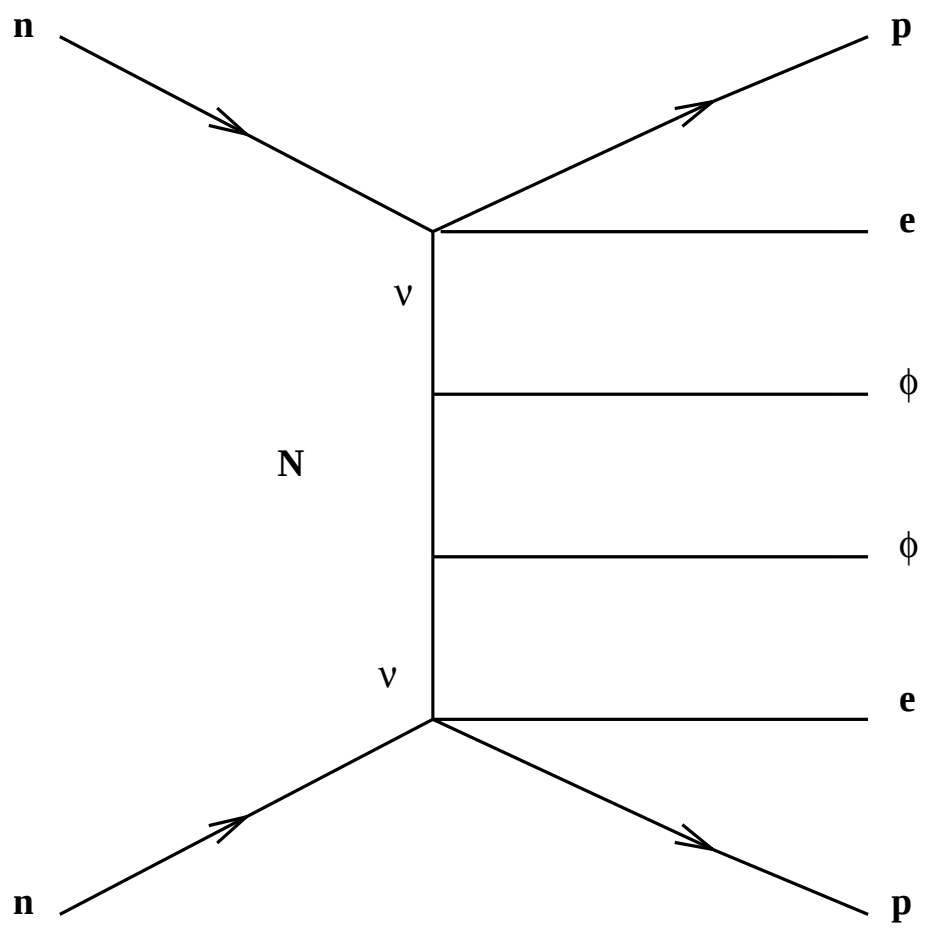

Fig. 2. Feynman graph for fermion–mediated $\beta\beta\phi\phi$–decays

With the nuclear matrix elements from[1,2] one can convert observed half lives (or limits thereof) into values for the effective Majoron neutrino coupling constant, according to:[6,14]

$$[T_{1/2}]^{-1} = |<g_\alpha>|^2 \cdot |M_\alpha|^2 \cdot G_{BB_\alpha} \tag{3}$$

for $\beta\beta\phi$-decays or

$$[T_{1/2}]^{-1} = |<g_\alpha>|^4 \cdot |M_\alpha|^2 \cdot G_{BB_\alpha} \tag{4}$$

for $\beta\beta\phi\phi$–decays. The index $\alpha$ in eqs. (3)–(4) indicates that effective coupling constants, matrix elements and phase spaces differ for different models.

The half lives of the different decay modes can be determined from the experimental spectra for the sum energy of the emitted electrons using the different spectral shapes (fig. 3) for the discrimination of the corresponding decay modes. For the evaluation a simultaneous maximum likelihood fit of the $2\nu\beta\beta$ decay and one selected Majoron–emitting decay has been performed.

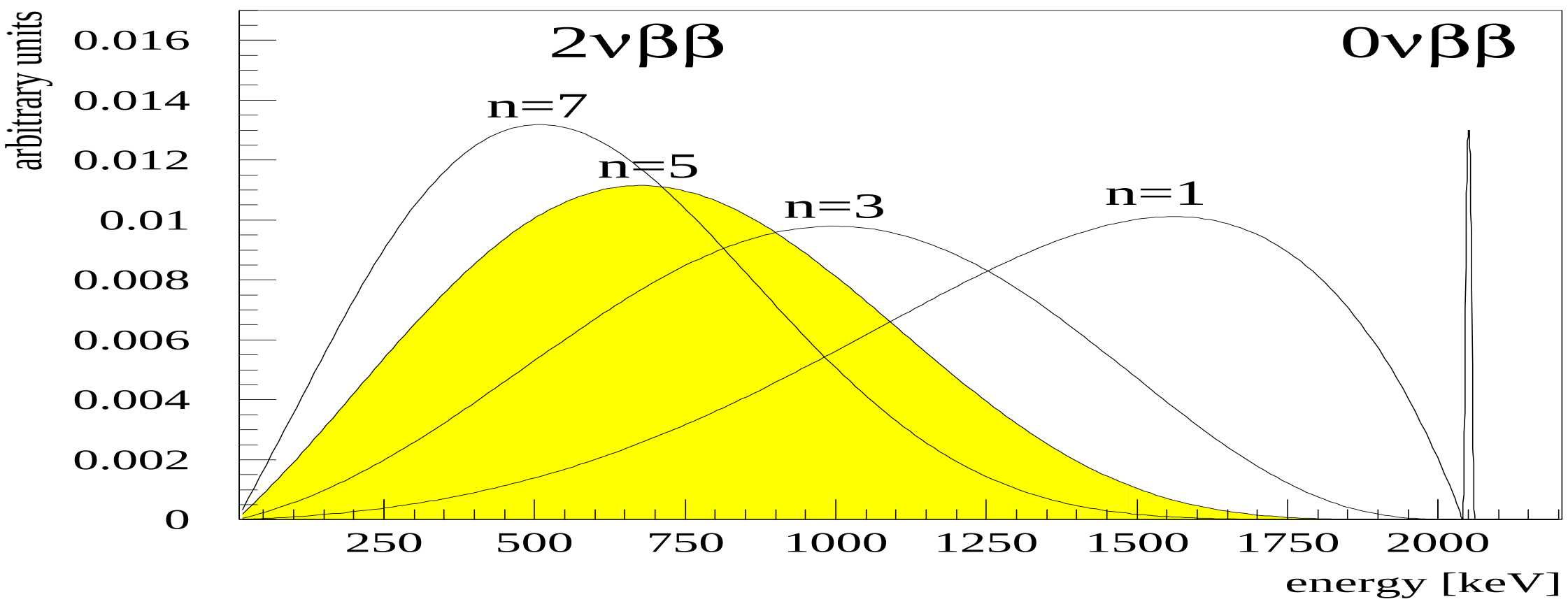

Fig. 3. Spectral shapes of the different decay modes

As input to the fit, the data taken with the enriched detector #2 of the Heidelberg–Moscow Double Beta Decay Experiment is used. This HP–Ge detector with an active mass of 2.758 kg is the biggest of the five with 86% in $^{76}$Ge enriched detectors operated in the Gran Sasso underground laboratory.[16,15] In the period between 9/1992 and 11/1994 the accumulated data with a measuring time of 640.962 d corresponds to a statistical significance of 4.84 kg·y.

Background due to natural radioactivity and other radioactive background sources has been subtracted prior to the fit. To unfold the background a Monte Carlo background model for the three detectors enr#1–enr#3 based on the CERN code GEANT3 was developed which is described explicitly in.[17] The measured activities in the setup are based on 47 identified $\gamma$–lines in the spectrum of the three enriched detectors enr#1–enr#3 and two separate activity measurements of $^{40}K$ and $^{210}Pb$ in the LC2–Pb. A uniform distribution of the activities inside a certain volume/material is assumed in the Monte–Carlo–simulation. The interaction and influence between each of the detectors activities with the neighboring detectors is fully included in the background model.

The data of enriched detector #2 was selected, because the raw data and the simulation show the best archieved low background of all detectors used in the Heidelberg–Moscow experiment.[17] The binning of the evaluated spectra is 20 keV per channel to avoid statistical fluctuations in the background model (the resolution of enr#2 is $2.43 \pm 0.02$ keV at 1332 keV) and the energy range of the fit has been chosen from 300 to 2040 keV (with the Q–value of $^{76}$Ge 2038.56 keV[18]). This range allows to maximize the available statistics, while minimizing the systematic errors of the background model, which increase drastically below 300 keV.[17] However all maxima of the different spectral shapes are included and choosing other energy ranges for the fit would not lead to significantly different results. In most cases with the above quoted energy range for the fit the most conservative limits are obtained.

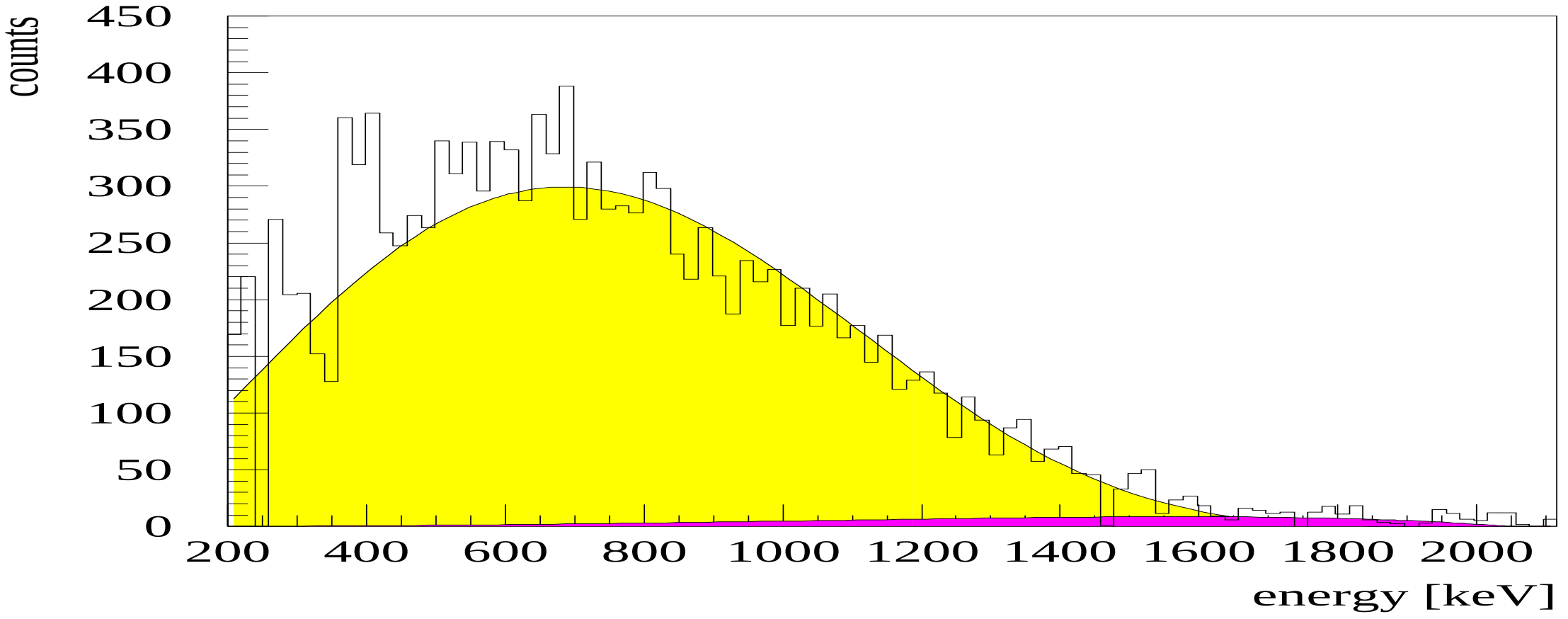

Fig. 4. "Ordinary Majoron" (n=1) in the area of fit: 300 – 2040 keV, yielding a halflife bound of $T_{1/2} > 7.91 \cdot 10^{21} y$ with $(90\% C.L.)$

The results of the data fits are shown in fig. 4 – 6. In each figure the experimental spectrum is shown as a histogram, while the light grey–shaded area is the best fit for the $2\nu\beta\beta$ decay. The dark shaded areas are the best fits for the different Majoron spectra, $n = 1$ in fig. 4, $n = 3$ in fig. 5 and $n = 7$ in fig. 6.

A clear discrimination of all Majoron emitting decays from the two neutrino emitting decay and consequently restrictive half life limits for the investigated decay modes are obtained.

The deviation from zero of Majoron emitting modes in the measured energy spectrum varies from $0.29\sigma$ (Ordinary Majorons) and $0.35\sigma$ (derivatively coupled $0\nu\beta\beta\phi\phi$) to $0.90\sigma$ (charged Majorons or non–derivatively coupled $0\nu\beta\beta\phi\phi$), meaning all effects of Majoron models are compatible with zero. The variation of the half life of the $2\nu\beta\beta$ decay in the fits to different Majoron emitting modes stays in a range between $1.67 \cdot 10^{21}$ years for a fit to $2\nu\beta\beta$ decay alone without any Majoron model and $1.86 \cdot 10^{21}$ years for charged or double ordinary Majoron decays. Also the evaluated half life for the $2\nu\beta\beta$ decay of $(1.77^{+0.13}_{-0.11}) \cdot 10^{21} y$ (68 % C.L.) in the evaluation interval 500–2040 keV[17] is in good agreement with these results.

Consequently only lower limits on Majoron emitting decay half lives are quoted. These are obtained by adding the statistical errors of the fits and the dominating systematical error of the background model in quadrature to the best fit half lives.

Restrictive limits on the coupling constants of ordinary Majoron models are found. Limits on any of the new Majoron models, however, are weaker by (3–4) orders

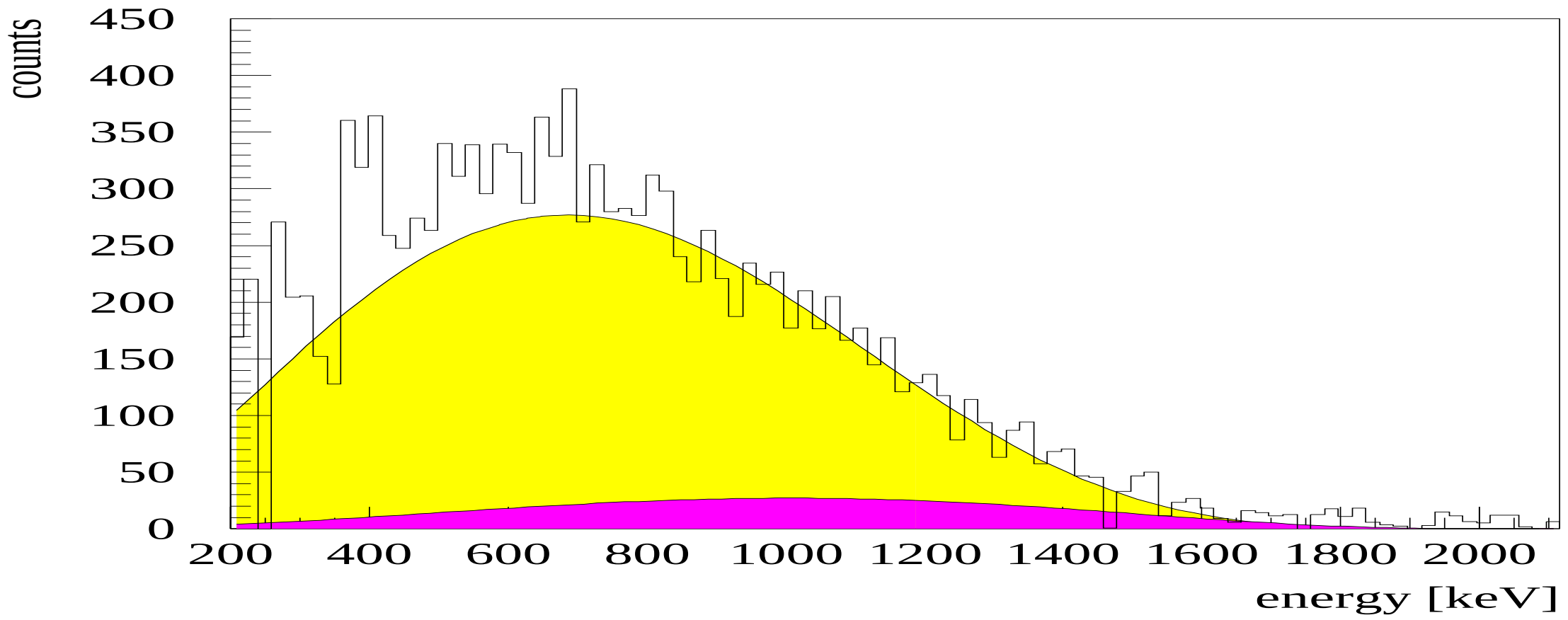

Fig. 5. "Charged Majoron" or "Double Majoron" (n=3) in the area of fit: 300 – 2040 keV, yielding a halflife bound of $T_{1/2} > 5.85 \cdot 10^{21} y$ with $(90\% C.L.)$

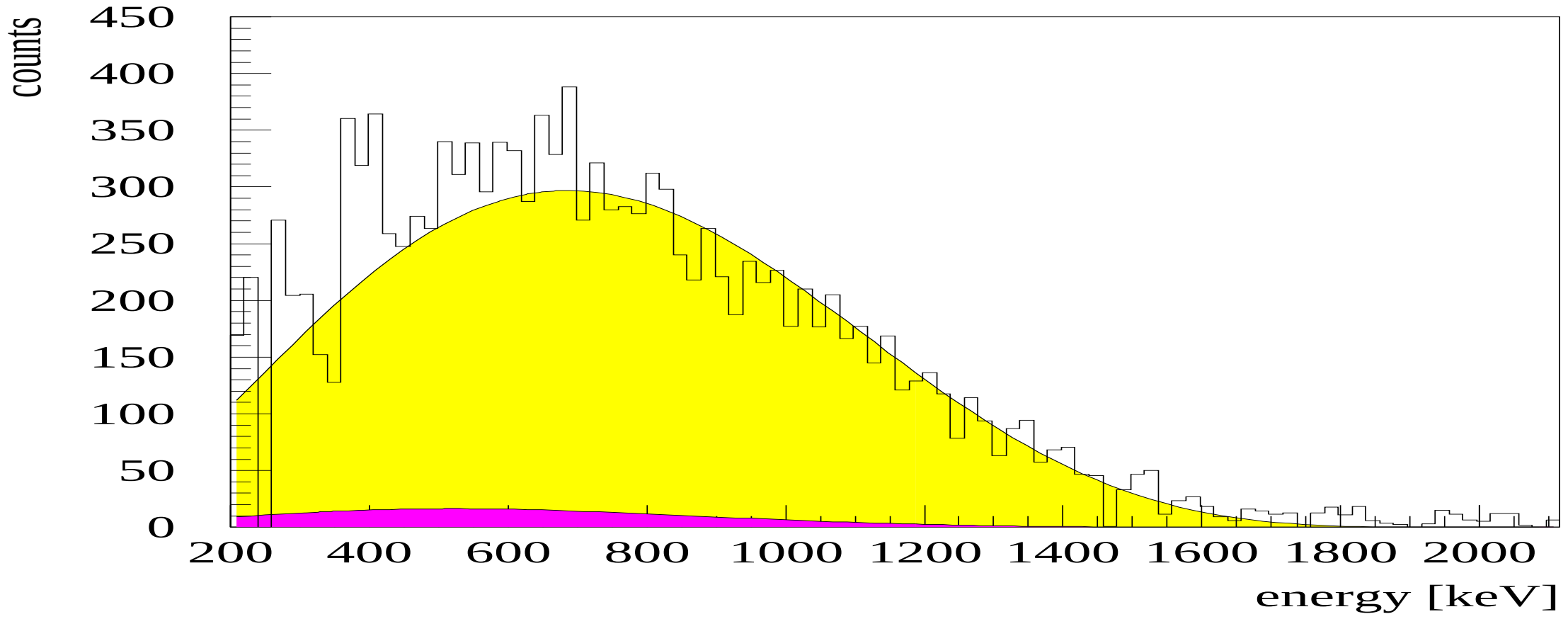

Fig. 6. "Double Majoron" (n=7) in the area of fit: 300 – 2040 keV, yielding a halflife bound of $T_{1/2} > 6.46 \cdot 10^{21} y$ with $(90\% C.L.)$

| case | modus | Goldstone boson | L | $T_{1/2} >$ (90%$C.L.$) | $g <$ (90%$C.L.$) |
|---|---|---|---|---|---|
| IB | $\beta\beta\phi$ | no GB | / | $7.91 \cdot 10^{21}$ | $2.3 \cdot 10^{-4}$ |
| IC | $\beta\beta\phi$ | GB | / | $7.91 \cdot 10^{21}$ | $2.3 \cdot 10^{-4}$ |
| ID | $\beta\beta\phi\phi$ | no GB | / | $5.85 \cdot 10^{21}$ | 4.1 |
| IE | $\beta\beta\phi\phi$ | GB | / | $5.85 \cdot 10^{21}$ | 4.1 |
| IIB | $\beta\beta\phi$ | no GB | -2 | $7.91 \cdot 10^{21}$ | $2.3 \cdot 10^{-4}$ |
| IIC | $\beta\beta\phi$ | GB | -2 | $5.85 \cdot 10^{21}$ | 0.18 |
| IID | $\beta\beta\phi\phi$ | no GB | -1 | $5.85 \cdot 10^{21}$ | 4.1 |
| IIE | $\beta\beta\phi\phi$ | GB | -1 | $6.64 \cdot 10^{21}$ | 3.3 |
| IIF | $\beta\beta\phi$ | Gauge boson | -2 | $5.85 \cdot 10^{21}$ | 0.18 |

Table 2. Bounds on half lives and coupling constants corresponding to the considered models deduced from the HEIDELBERG–MOSCOW experiment

of magnitude, although the experimental half life limits are comparable for all decay modes!

Note that the surprisingly weak limits obtained for all of the new Majoron models are caused by the small values of the corresponding nuclear matrix elements and phase spaces and is independent of the isotope under consideration. Similarly weak limits will be obtained by any double beta decay experiment with comparative sensitivity in the half life limits.[1,2]

In summary, motivated by recent theoretical work on Majorons[10–12] an analysis of our experimental data has been carried out to derive limits on the half lives of the various Majoron emitting decay modes for $^{76}Ge$. Combining these results with the nuclear matrix elements listed in[1,2] limits on the Majoron–neutrino coupling were derived, for several cases for the first time.

Restrictive limits on the half life of all Majoron emitting decay modes (meaning a clear discrimination of the different spectral shapes) have been obtained. For the effective coupling constants of the classical Majoron models predicting $\beta\beta\phi$–decays stringent limits have been obtained. However, for the new Majoron models, much weaker limits have been obtained due to the small values of nuclear matrix elements and phase space integrals.

**Acknowledgements**

The HEIDELBERG–MOSCOW experiment is supported by the Bundesministerium für Forschung und Technologie der Bundesrepublik Deutschland and the Ministry of Science and Technology of Russian Federation. M.H. is supported by the Deutsche Forschungsgemeinschaft (446 JAP–113/101/0 and Kl 253/8–1). B.M. is supported by a Human Capital and Mobility fellowship of the European Community (ERBCHBGCT928183).

**References**

1. M. Hirsch *et al.*, to be appear